\documentclass[aps,prx,onecolumn,superscriptaddress,longbibliography]{revtex4-2}

\usepackage{amsmath,amsfonts,qcircuit,bbm}
\usepackage[colorlinks=true,breaklinks=true,linkcolor=magenta,citecolor=magenta,urlcolor=magenta]{hyperref}
\usepackage{graphicx}

\usepackage{listings}
\usepackage{parskip}
\usepackage{braket}
\usepackage{makecell}
\usepackage{tabularx}
\usepackage[table]{xcolor}
\usepackage{verbatim}

\newcommand{\vect}[1]{{\bf{#1}}}

\begin{document}

\title{A quantum computational approach to the open-pit mining problem}

\author{Yousef Hindy}
\thanks{These authors contributed equally to this work.}
\affiliation{Department of Computer Science, Stanford University, 353 Jane Stanford Way, Stanford, CA 94305, USA}
\author{Jessica Pointing} 
\thanks{These authors contributed equally to this work.}
\affiliation{Department of Computer Science, Stanford University, 353 Jane Stanford Way, Stanford, CA 94305, USA}
\affiliation{Department of Physics, University of Oxford, Oxford OX1 2JD, United Kingdom}
\author{Meltem Tolunay}
\thanks{These authors contributed equally to this work.}
\affiliation{Department of Electrical Engineering, Stanford University, 350 Jane Stanford Way, Stanford, CA 94305, USA}
\author{Sreeram Venkatarao}
\thanks{These authors contributed equally to this work.}
\affiliation{Department of Computer Science, Stanford University, 353 Jane Stanford Way, Stanford, CA 94305, USA}
\author{Mario Motta}
\affiliation{IBM Quantum, IBM Research Almaden, 650 Harry Road, San Jose, CA 95120, USA}
\author{Joseph A. Latone}
\thanks{corresponding author, e-mail: jlatone@us.ibm.com}
\affiliation{IBM Quantum, IBM Research Almaden, 650 Harry Road, San Jose, CA 95120, USA}

\begin{abstract}
The determination of optimal open-pit profiles is 
a well-studied combinatorial optimization problem, 
with profound technical and conceptual relevance in computational mining.
The ongoing evolution of quantum computing hardware 
and the recent advances of heuristic quantum algorithms 
make it worthwhile to explore 
the solution of the open-pit mining problem on quantum computers.
In this work, we cast the open-pit mining problem 
as a Hamiltonian ground-state search problem, 
which in turn we solve with a dedicated implementation 
of the variational quantum eigensolver algorithm, 
and we propose a domain decomposition approach 
to extend the reach of today's small scale quantum hardware.
The procedure is demonstrated on IBMQ devices using four qubits.
This is the first example, to the best of our knowledge, 
of open-pit profile calculations being performed on quantum hardware.
\end{abstract}

\maketitle

\section{Introduction}

Combinatorial optimization problems have long been recognized 
as applications for a quantum computer \cite{farhi2014quantum,montanaro2016quantum,otterbach2017unsupervised,moll2018quantum,neven2008image}.
Solving a combinatorial optimization problem typically means 
finding the global maximum of a function $C(\vect{z})$ of $n$ 
binary variables $\vect{z} = (z_0 \dots z_{n-1})$, $z_i \in \{0,1\}$.
The exponential growth of the space of length-n binary strings 
with $n$ makes brute-force search algorithms intractable.
This limitation motivated the design of heuristic algorithms 
for classical computers that find local maxima of the cost 
function with polynomial cost, for example leveraging notions 
of graph theory \cite{hochbaum2008pseudoflow,lucas2014ising}.

More recently, quantum algorithms were proposed as an alternative and 
complementary route to the solution of combinatorial optimization problems
\cite{farhi2014quantum,montanaro2016quantum,moll2018quantum}.
It is not yet known if a quantum algorithm can provide computational 
advantage in a combinatorial optimization problem, especially with the 
limitations of contemporary hardware \cite{bennett1997strengths,bernstein1997quantum,crooks2018performance,guerreschi2019qaoa,elfving2020will,zhou2020quantum}. 
Designing quantum heuristics is nevertheless a valuable avenue of research, 
as they could be practically useful for finding near-optimal solutions. 

Adiabatic state preparation (ASP) 
\cite{farhi2000quantum,farhi2001quantum}
offers a polynomial cost solution to combinatorial optimization problems
under suitable conditions specifying a precise domain of applicability.
The method often requires a circuit depth and a number of entangling gates 
far exceeding the capabilities of contemporary quantum hardware.
These limitations motivated the development of heuristic quantum algorithms 
trading exactness with more moderate use of quantum resources.
Notable examples are the quantum approximate optimization algorithm (QAOA) 
\cite{farhi2014quantum,hadfield2019quantum},
which is a variational quantum algorithm with ansatz inspired by ASP, 
the variational quantum eigensolver (VQE)
\cite{peruzzo2014variational,kandala2017hardware,nannicini2019performance},
where the variational principle is combined with an arbitrarily flexible 
and hardware-aware ansatz.
Other heuristic algorithms have been proposed as well
\cite{motta2020determining,mcardle2019variational}.

In this work, we explored a combinatorial optimization problem motivated by 
open-pit mine design \cite{lerchs1965optimum,johnson1968optimum,bienstock2010solving,hochbaum2000improved,meagher2014optimized,brazil2015optimal,lagos2020framework,hochbaum2000performance,lamghari2014variable,lamghari2020hyper}.
The definition of a open pit mine is "an excavation or cut made at the surface 
of the ground for the purpose of extracting ore and which is open to the 
surface for the duration of the mine's life" \cite{fourie1992open}.
To expose and mine the ore, it is generally necessary to excavate and relocate 
large quantities of waste rock.
The main objective in any commercial mining operation is the exploitation 
of the mineral deposit at the lowest possible cost,
under constraints imposed by geologic and mining engineering aspects, 
e.g. environmental impact and structural safety.
An open-pit mining operation can be essentially viewed as a process by which 
the open surface of a mine is continuously deformed, and the planning of a 
mining program involves the design of the final shape of this open surface.

Following published literature in the field \cite{lerchs1965optimum}, 
we will assume that the type of material, its mining value and its extraction 
cost are given for each point of the mine, and that restrictions on the geometry 
of the pit are specified (surface boundaries and maximum allowable wall slopes).
The combinatorial optimization problem tackled here is to maximize the total 
profit of the excavation process, defined as the total mine value of material 
extracted minus total extraction cost.
Published literature indicates that the open-pit profile design problem, 
when formulated in three spatial dimensions and in presence of constraints 
dictated by the nature of the mining and processing operations,
is in the NP-hard complexity class \cite{bienstock2010solving,lamghari2014variable}.
On the other hand, if the mining and processing constraints are eliminated, 
and if the scheduling horizon reduces to a single period, then the open-pit 
profile design problem reduces to an instance of the minimum-cut problem \cite{picard1976maximal}, which is solvable at polynomial cost on a classical 
computer \cite{lamghari2014variable,lamghari2020hyper}.

In the open-pit mining literature, several classical algorithms have been proposed, which include heuristic approximation, genetic and network flow algorithms \cite{lerchs1965optimum,fourie1992open,picard1976maximal,hochbaumchen2000,pseudoflow,caccetta1986,thomas1996pit}. One of the most widely recognized solutions to the open-pit problem is the Lerchs-Grossmann (LG) algorithm \cite{lerchs1965optimum,bond1995mathematical}, which uses a graph theoretic approach and solves the open-pit optimization problem by converting it into a maximum closure problem. After the development of the LG method, other network flow algorithms were shown to compute the same results faster. Among those, the pseudoflow algorithm \cite{pseudoflow} is one of the most widely accepted methods for the open-pit optimization problem. It is also important to note that these algorithms are heuristic and rely on mathematical approximations that scale well with the size of an open-pit problem, but do not guarantee an exact solution. In our work, we make a first attempt to map the open-pit problem to a quantum setting, and to examine the potential of quantum algorithms to deliver heuristic solutions.

\section{Methods}
\subsection{Hamiltonian formulation}

To provide an approximate description of an open-pit mine, we model a portion of ore by a collection of finite volume elements (or blocks)
with coordinates $i = (x_i,y_i)$ in 2D or $i = (x_i,y_i,z_i)$ in 3D, forming a lattice $\mathrm{L}$ of $n$ sites.
We assume that the economic value $v_{i}$ and digging cost $c_{i}$ of each block is known, 
which in turn determine the profit $w_{i} = v_{i} - c_{i}$ from digging that block.
To every block $i$ we also associate a binary number $z_{i}$ such that $z_{i}=1$ (0) if that block is excavated (not excavated).
Our goal is to maximize the profit $P(\vect{z}) = \sum_i w_i \, z_i$ under a constraint dictated by the open nature of the pit: every excavated block must be connected with the surface,
and the walls of the pit must not exceed a maximum steepness.
We refer to these conditions as "smoothness constraint"
throughout the remainder of the present work.

\begin{figure}[b!]
\includegraphics[width=0.8\textwidth]{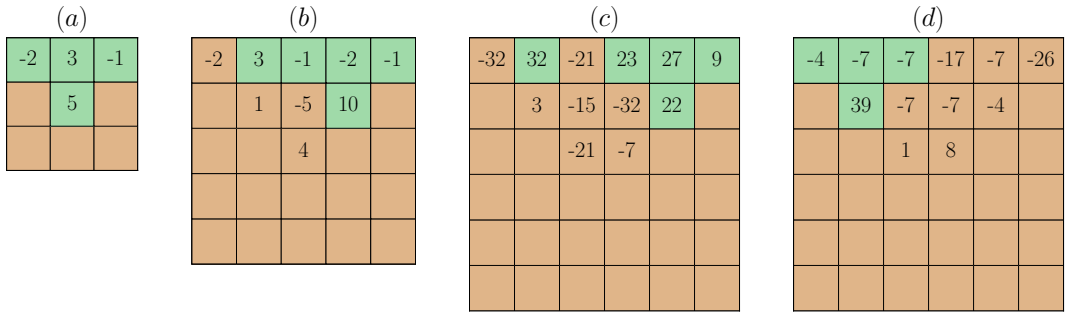}
\caption{Pit profiles studied in the present work. On each block of ore (square cells) the profit $w_{i}$ is shown. Green (brown) blocks correspond to excavated (unexcavated) blocks in an optimal solution of the open pit problem, and maximum profits are 5, 9, 113 and 21 respectively. The pit profiles are represented with systems of 4, 9, 12, and 12 qubits respectively.}
\label{fig:pit}
\end{figure}

To enforce the smoothness constraint we require that, if a block $i$ is excavated, so must be all the blocks lying above it. We call these blocks the "parents" of $i$ and denote them as $j \in P_i$. If the block $i$ has coordinates $(x_i,y_i,z_i)$, its parents have coordinates $z_j = z_i + 1$ and $x_j - x_i \leq 1$, $y_j - y_i \leq 1$.
The smoothness constraint is then mathematically expressed requiring that the following smoothness function $S(\vect{z}) = \sum_i \sum_{j \in P_i} z_i (1-z_j)$ vanishes. Examples of pit profiles satisfying the
smoothness constraint are shown in Figure \ref{fig:pit}: every excavated block (colored green) is connected to the surface, and the walls of the pit (set of blocks colored green) have slope below 45 degrees.

The open-pit design problem is very naturally mapped onto a qubit problem.
To each block $i$ we associate a qubit, so that the binary string $\vect{z} = (z_0 \dots z_{n-1})$ corresponds to the state 
$|\vect{z} \rangle = \otimes_{i=0}^{n-1} | z_ i \rangle$. The profit and smoothness functions are mapped onto the operators
\begin{equation}
\label{eq:vqe_ansatz}
\hat{H}_p = \sum_i w_i \, \frac{1-Z_i}{2} 
\quad,\quad
\hat{H}_s = \sum_i \sum_{j \in P_i} \frac{1-Z_i}{2} \otimes \frac{1+Z_j}{2} 
\quad.
\end{equation}
It is worth remarking that $\hat{H}_s$ is non-negative, and its null space is the span of states $|\vect{z} \rangle$ such that $S(\vect{z}) = 0$.

In this framework, the open-pit design problem thus corresponds to finding binary strings $\vect{z}$ that lie in the ground eigenspace of $\hat{H}_p$ 
and in the null space of $\hat{H}_s$ (optimal binary strings).
Here, we replace this constraint problem by an unconstrained problem formed adding a term to the objective function $\hat{H}_p$,
that consists of a penalty parameter $\gamma > 0$ multiplied by the operator $\hat{H}_s$, taken as a measure of violation of the smoothness constraint.
Therefore, we search for optimal binary strings by minimizing the expectation value of the Hamiltonian $\hat{H} = \hat{H}_p + \gamma \, \hat{H}_s$.
In this work, we explore the use of the variational quantum eigensolver, and we propose the following ansatz,
\begin{equation}
\label{eq:ansatz}
| \Psi(\theta) \rangle = \prod_{i} \prod_{j \in  P_i} c R_y(\theta_{ij}) \prod_i R_y(\theta_{i}) | \Psi_0 \rangle
\;,
\end{equation}

where a register of qubits is prepared in a tensor product $| \Psi_0 \rangle$ of single-qubit states (here $| 0 \rangle^{\otimes n}$, $| 1 \rangle^{\otimes n}$ or
$|+\rangle^{\otimes n}$) and manipulated with a sequence of single-qubit $Y$ rotations ($R_y(\theta_{i})$, applied to qubit $i$) and a sequence of controlled-$Y$ rotations ($c R_y(\theta_{ij})$, applied to qubits $i$ and $j$, where the target qubit $j$ is a parent of the control qubit $i$).
The ansatz \eqref{eq:ansatz} balances two needs: the desire to 
keep a limited circuit depth, and the desire to connect the ansatz 
structure with the digging process. Single-qubit $Y$ rotations continuously change the state of a qubit between excavated and unexcavated (0 and 1 respectively), allowing to explore various
binary strings ${\bf{z}}$. Controlled-$Y$ rotations excavate the children $j$ of a parent $i$, if the qubit corresponding to $i$ is in state $0$ (excavated), and thus help enforcing smoothness constraints.
$Y$-rotations are also chosen to maintain the wavefunction 
real-valued as the optimization unfold, thereby reducing the 
number of variational parameters.
We remark that the choice \eqref{eq:ansatz} is not unique, 
and the exploration of other ansatze is a valuable effort.

\subsection{Domain decomposition}

As contemporary quantum devices are limited in the number and quality of qubits, they currently cannot tackle many problems of practical size directly. 
Variational algorithms such as VQE address the challenge posed by the quality of qubits, 
as they bypass the issue of quickly accumulating errors by imposing the use of shallow quantum circuits comprising a limited number of gates.
A natural way to address the challenge posed by the number of qubits is to decompose the problem into a collection of subproblems,
solve these subproblems on a small quantum device and combine these solutions on a classical computer to obtain a global solution
\cite{bian2016mapping,rosenberg2016building,karimi2017effective,shaydulin2018community,shaydulin2019network,shaydulin2019hybrid}.
Here, we partition the lattice $\mathcal{L}$ into a collection $\{ F_\alpha \}_\alpha$ of mutually disjoint fragments $F_\alpha$, each comprising a group of blocks.
Correspondingly, the Hamiltonian is written (see Appendix \ref{sec:frag}) as 
\begin{equation}
\label{eq:frag1}
\hat{H} = \sum_\alpha \hat{V}_\alpha + \frac{1}{2} \, \sum_{\beta \neq \alpha} \hat{W}_{\alpha,\beta}
\quad.
\end{equation}
We focus on wavefunctions of the form $| \Psi \rangle = \otimes_\alpha | \Phi_\alpha \rangle$,
where $| \Phi_\alpha \rangle$ has support on the qubits associated with fragment $F_\alpha$.
The expectation value of $\hat{H}$ over a factored wavefunction is minimized when the 
wavefunctions $| \Phi_\alpha \rangle$ describing individual fragments satisfy a stationary Schr\"odinger equation of the form
\begin{equation}
\label{eq:frag2}
\hat{T}_\alpha [ \Phi ] | \Phi_\alpha \rangle = E_\alpha | \Phi_\alpha \rangle 
\quad,\quad
\hat{T}_\alpha [ \Phi ] = \hat{V}_\alpha + 
\sum_\beta \mbox{Tr}[ \mathbbm{I} \otimes | \Phi_\beta \rangle \langle \Phi_\beta | \hat{W}_{\alpha,\beta}] 
+ 
\sum_\beta \mbox{Tr}[ | \Phi_\beta \rangle \langle \Phi_\beta | \otimes \mathbbm{I} \hat{W}_{\beta,\alpha}] 
\quad,
\end{equation}
which we solve self-consistently.

\section{Results}

In this work, we consider the instances of the open pit mining problem shown in 
Figure \ref{fig:pit}. Such instances are chosen to test the proposed algorithm 
on small instances of different regimes of operation. 
Profile $(a)$ is a minimal example, where all sites are excavated.
The slightly larger profile $(b)$ illustrates the effect of the smoothness constraint 
(excavation of the blocks with profits $w_i=-1$). 
In profile $(c)$, profit is concentrated along a diagonal ($w_i = 23,22$), 
providing a minimal mathematical representation of a stringer of valuable material, 
and in profile $(d)$ the profit varies with more continuity across the pit profile.

To carry out numerical simulations, we used IBM's open-source Python library for 
quantum computing, Qiskit \cite{aleksandrowicz2019qiskit}. 
Qiskit provides tools for various tasks such as creating quantum circuits, 
performing simulations, and computations on real hardware. 
It also contains an implementation of the VQE algorithm, 
a hybrid quantum-classical algorithm that uses both quantum and classical resources 
to solve the Schr\"{o}dinger equation and a classical exact eigensolver algorithm 
to compare results.
In the VQE algorithm, we take our wavefunction in the form of a quantum circuit, 
which is of the type described in Equation~\eqref{eq:vqe_ansatz}.
We then minimize the expectation value of the Hamiltonian with respect to the 
parameters of our circuit. The minimization is carried out through the classical 
optimization method L-BFGS-B \cite{zhu1997algorithm,byrd1995limited,morales2011remark} 
on the simulator, and Simultaneous Perturbation Stochastic Approximation (SPSA) \cite{spall1998overview,spall2000adaptive} on quantum hardware. 
The performance of different optimizers is compared in Appendix \ref{sec:app_opt}.
Once the VQE is complete, we obtain the optimized variational form and the estimate 
for the ground state energy. 

\begin{figure}[b!]
\includegraphics[width=0.3\textwidth]{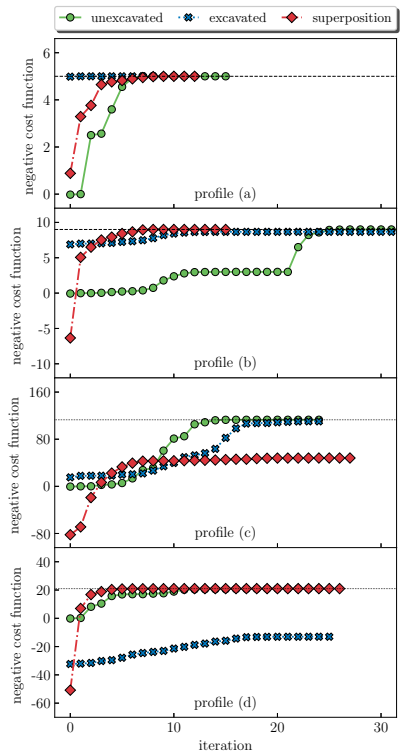}
\caption{Evolution of the VQE cost function in the simulation of profiles 
in Figure \ref{fig:pit} (top to bottom), using different initial wavefunctions 
$| \Psi_0 \rangle = | 0 \rangle^{\otimes n}$, $| 1 \rangle^{\otimes n}$ 
and $| + \rangle^{\otimes n}$ (respectively denoted unexcavated/excavated/superposition 
and shown in green/blue/red). Black dashed lines denote maximum profit in 
optimal solutions of the problem.}
\label{fig:vqe_sv_large_mines}
\end{figure}

In addition, we measure the qubits in the register, 
obtaining probabilities for various binary strings,
\begin{equation}
p({\bf{z}}) = | \langle \Psi(\theta) | {\bf{z}} \rangle |^2
\;.
\end{equation}
From these, we compute the probability to extract a binary string 
corresponding to an optimal solution of the open pit design problem,
\begin{equation}
p_{\mathrm{opt}} = \sum_{\substack{ {\bf{z}}: S({\bf{z}})=0, \\ P({\bf{z}}) = P_{\mathrm{opt}}}} p({\bf{z}}) 
\quad,\quad
P_{\mathrm{opt}} = \max_{{\bf{z}}} P({\bf{z}})
\quad.
\end{equation}
We performed hardware experiments on the $\mathsf{ibmq\_paris}$ quantum device. 
We employed readout-error mitigation 
\cite{temme2017error,kandala2019error,bravyi2020mitigating} as implemented in 
the Qiskit Ignis library, to correct measurement errors. 

\subsection{Simulations on classical computer}

In Figure \ref{fig:vqe_sv_large_mines} and Table \ref{tab:1} we demonstrate 
convergence of the VQE algorithm for the three largest profiles in Figure \ref{fig:pit}, 
using initial wavefunctions $| \Psi_0 \rangle = | 0 \rangle^{\otimes n}$, 
$| 1 \rangle^{\otimes n}$ and $| + \rangle^{\otimes n}$ (excavated, unexcavated, 
superposition) and initial VQE parameters distributed uniformly in the interval 
$(-\pi/10,\pi/10)$.
The optimization is generally more challenging for profiles $(c)$ and $(d)$, where 
the larger system size and the continuous nature of the distribution of profit makes 
more difficult to identify and excavate valuable portions of ore. Convergence is 
typically more rapid and systematic for initial state $| 0 \rangle^{\otimes n}$. 
As seen in Table \ref{tab:1}, when the conditions $\langle \hat{H}_s \rangle = 0$, 
$\langle H_{p} \rangle = \max_{\bf{z}} P({\bf{z}})$ for successful convergence are
met, measurement of the qubits yields the optimal pit profile with probability 1.

\begin{table}[t!]
\begin{tabular}{cccccccccccccccccccc}
\hline\hline
profile & $|\Psi_0\rangle$ & $\gamma$ & $p_{\mathrm{opt}}$ & &
profile & $|\Psi_0\rangle$ & $\gamma$ & $p_{\mathrm{opt}}$ & &
profile & $|\Psi_0\rangle$ & $\gamma$ & $p_{\mathrm{opt}}$ & &
profile & $|\Psi_0\rangle$ & $\gamma$ & $p_{\mathrm{opt}}$ \\
\hline
a & $| 0 \rangle^{\otimes n}$ &  7/3 & 1.00 & & 
b & $| 0 \rangle^{\otimes n}$ & 10/3 & 1.00 & & 
c & $| 0 \rangle^{\otimes n}$ & 53/3 & 1.00 & & 
d & $| 0 \rangle^{\otimes n}$ & 22/3 & 1.00 \\
a & $| 1 \rangle^{\otimes n}$ & 7 /3 & 1.00 & & 
b & $| 1 \rangle^{\otimes n}$ & 10/3 & 1.00 & & 
c & $| 1 \rangle^{\otimes n}$ & 53/3 & 1.00 & & 
d & $| 1 \rangle^{\otimes n}$ & 22/3 & 0.00 \\
a & $| + \rangle^{\otimes n}$ &  7/3 & 1.00 & & 
b & $| + \rangle^{\otimes n}$ & 10/3 & 1.00 & & 
c & $| + \rangle^{\otimes n}$ & 53/3 & 0.00 & & 
d & $| + \rangle^{\otimes n}$ & 22/3 & 1.00 \\
\hline\hline
\end{tabular}
\caption{Penalty parameter $\gamma$ and probability
of detecting an optimal solution of the pit profile problem $p_{\mathrm{opt}}$ 
for the calculations reported in Figure \ref{fig:vqe_sv_large_mines}.
Penalty parameters were initialized to $\gamma_0 = \max_i ( w_i - \sum_{j \in P_i} w_j )/3$ and then varied manually by integer multiples of $1$ until maximization of $p_{opt}$ was achieved.}
\label{tab:1}
\end{table}

Further insight into the convergence of VQE is offered by the analysis of the expectation values of the single-qubit Pauli $Z$ operators, $E[Z_i] = \langle \Psi(\theta) | Z_i | \Psi(\theta) \rangle$, which in turn provide the probability distribution $p(z_i = k) = (1+(-1)^k E[Z_i])/2$, $k=0,1$, for the measurement of a single qubit, and offer a way of visualizing the pit profile as the optimization of the VQE cost function unfolds.

\begin{figure}[t!]
\includegraphics[width=0.5\columnwidth]{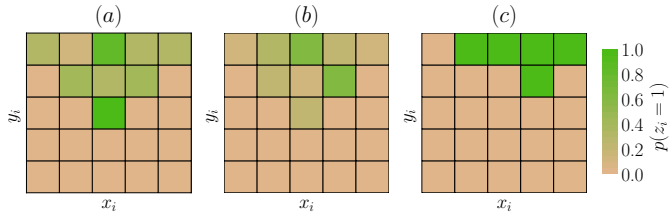}
\caption{Evolution of the mining profile, i.e. the probability $p(z_i = 1)$
for each site $i$, at three different moments (marked a,b,c) of the optimization. 
Green (brown) denote excavated (unexcavated) sites.}
\label{fig:nine_pit_mine_probability}
\end{figure}

In Figure \ref{fig:nine_pit_mine_probability}, we show the probabilities $p(z_i = 1)$, for the pit profile in Figure \ref{fig:pit}b, at three steps of the optimization. As seen, at the beginning of the optimization, $p(z_i = 1) < 1$ for all sites. On the other hand, at the end of the optimization, all sites have either $p(z_i = 1) = 1$ or $0$, signaling convergence of VQE to an element of the computational basis $|{\bf{z}} \rangle$ corresponding to the optimal configuration in Figure \ref{fig:pit}b.

\subsection{Domain decomposition}

\begin{figure}[b!]
\includegraphics[width=0.7\textwidth]{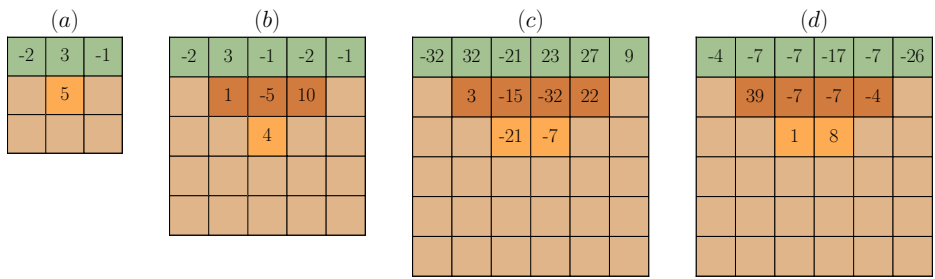}
\caption{\textit{Horizontal} cuts considered in fragmentation simulations.}
\label{fig:pit_h}
\end{figure}

\begin{figure}[t!]
\includegraphics[width=0.8\textwidth]{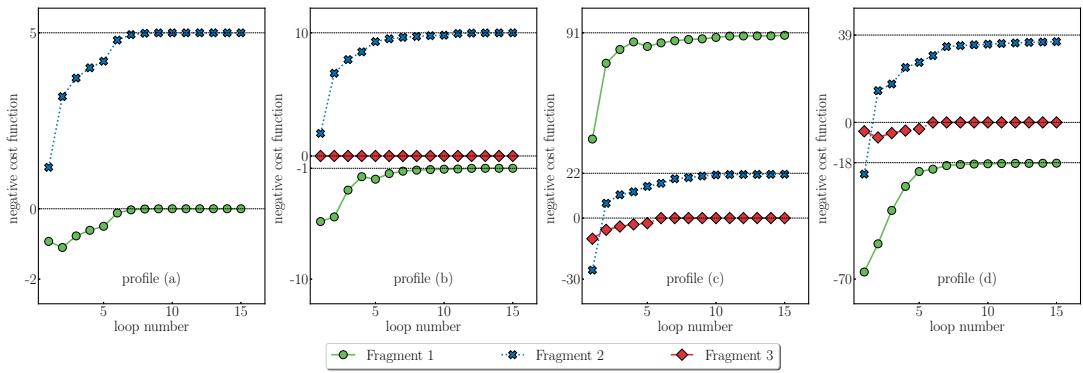}
\caption{Evolution of the negative loss function of the \textit{horizontal} fragments depicted in Figure \ref{fig:pit_h} using domain decomposition.}
\label{fig:fragmentation_2}
\end{figure}

In this Section, we explore the performance of domain decomposition. 
An illustrative calculation is reported in Figure \ref{fig:fragmentation_2}.  
We divide all pit profiles into fragments corresponding to horizontal stripes 
as depicted in Figure \ref{fig:pit_h}.
In each loop over all fragments, one iteration per fragment is performed. 
All fragments converge to the sum of their excavated sites in the optimal solution, 
thus the sum of the negative cost function of all fragments converges to the 
maximum possible profit of that pit profile, which is 5, 9, 113 and 21 respectively. 
All simulations are run using statevector simulator with L-BFGS-B optimizer, 
and penalties are set to 4, 5, 10 and 20 respectively. 
Initial states are all set to equal superposition.
It is worth emphasizing that the performance of fragmentation is influenced by 
the choice of the partition in fragments, and on the quality of the initial 
collection of fragment wavefunctions $| \Phi_\alpha \rangle$. These effects are 
documented in Appendix \ref{sec:frag_comparison}.

\subsection{Hardware experiments}

\begin{figure}[b!]
\includegraphics[width=0.4\columnwidth]{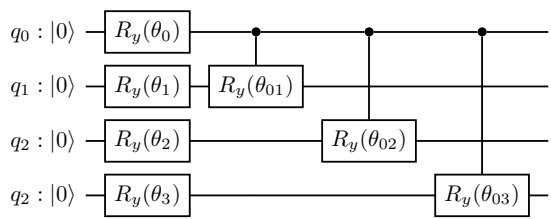}
\caption{
Quantum circuit used to study the 4-site pit
profile shown in Figure \ref{fig:pit}a.
}
\label{fig:hardware_0}
\end{figure}

We conclude by studying the profile in Figure~\ref{fig:pit}a on quantum hardware.
To study this 4-site pit profile, we use quantum circuit shown in Figure
\ref{fig:hardware_0}.
Simulations are carried out on the $\mathsf{IBMQ\_paris}$ quantum device, 
using qubits [22,24,26,25] for sites 0,1,2,3 respectively. This choice is motivated 
by the desire of matching qubit connectivity with parent relationship, to avoid 
an overhead of swap gates in the implementation of quantum gates on non-adjacent qubits.

\begin{figure}[t!]
\includegraphics[width=0.4\columnwidth]{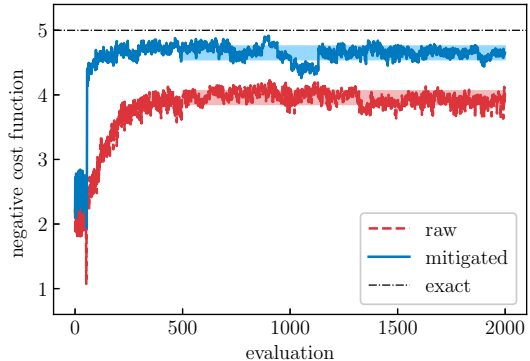}
\caption{
Evolution of the negative cost function, i.e. the expectation value of the operator 
$-\hat{H}$, during the variational optimization. Red, blue symbols denote
raw and readout-error-mitigated results respectively, and the dashed horizontal 
black line denotes the optimal negative cost function value. Colored bands denote a 67\%
confidence interval for the negative cost function samples after convergence 
of the optimization.}
\label{fig:hardware_1}
\end{figure}

\begin{figure}[t!]
\includegraphics[width=0.45\columnwidth]{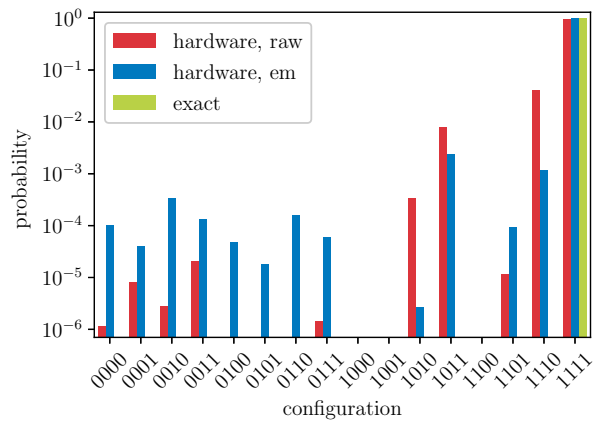}
\caption{
Probability distribution for the measurement of the four qubits of the register
in Figure \ref{fig:hardware_0} over the converged VQE wavefunction using a 
noiseless classical simulator of a quantum computer (green) and on quantum 
hardware with (blue) and without (red) readout error mitigation.
In all cases, the optimal open-pit configuration $(1,1,1,1)$ is the most likely 
outcome of the measurement.
}
\label{fig:hardware_2}
\end{figure}

The evolution of the cost function in the VQE calculation in shown in Figure 
\ref{fig:hardware_1}. We use the SPSA optimizer, which is a stochastic optimizer
particularly well-suited for simulations on quantum hardware, and employ readout
error mitigation.
The average cost function is $C_{raw} = 3.956 \pm 0.003$ and 
$C_{mitigated} = 4.651 \pm 0.003$, indicating that readout error mitigation reduces 
deviations between exact and VQE results by 66\% for this particular problem.

In Figure \ref{fig:hardware_2} we show the probability distribution for the measurement 
of the four qubits for the raw and error-mitigated hardware simulations, and for the 
exact ground state. The probability to obtain an optimal pit profile when measuring the 
qubits is $p_{\mathrm{opt}} = 0.951$, $0.996$ and $1$ in the three cases,
and in all cases the probability of measuring a profile violating smoothness
conditions is $p_{\mathrm{v}} < 0.01$.
The Batthacharyya distances between the three probability distributions are
\begin{equation}
d(\mathrm{raw},\mathrm{exact}) = 0.025 \quad,\quad 
d(\mathrm{mitigated},\mathrm{exact}) = 0.002 \quad,\quad 
d(\mathrm{raw},\mathrm{mitigated}) = 0.027 \quad,
\end{equation}
a further confirmation of the good quality of our simulations, and of the effectiveness 
of readout error mitigation techniques.

\section{Conclusions}

In this work, we proposed to tackle the problem of computational optimization of open-pit 
profiles by quantum computational algorithms.
We formulated the problem as a quadratic constrained binary combinatorial optimization 
problem, introducing a Hamiltonian operator that comprises two terms: $\hat{H}_p$, 
aimed at maximizing the profit of excavating a certain pit profile, and $\hat{H}_s$, 
aimed at enforcing continuity of the pit profile.
We introduced a domain decomposition approach to simplify and accelerate the quantum 
computational treatment of the open-pit profile optimization problem,
which can produce results of quality comparable with an original calculation, albeit 
tackling a collection of smaller and simpler problems. We discussed the limitations 
of such a domain decomposition approach, namely the sensitivity to initial conditions 
and partition into fragments. The latter difficulty can be alleviated by employing 
fragments of increasingly large size.

We elected to approximate the ground state of $\hat{H}_p + \hat{H}_s$ using the VQE 
algorithm due to its widespread use in published literature and computational packages. 
Nevertheless, we introduced a tailored ansatz comprising $y$ rotations and 
controlled-$y$ rotations, which proved able to yield accurate approximations for the 
pit profiles considered here. It is worth pointing out that, for larger and more 
complex problems, such an ansatz may yield inaccurate results \cite{mcclean2018barren} 
and improved ansatze may have to be designed, on the basis of mathematical 
considerations and empirical data \cite{grimsley2019adaptive,mccaskey2019quantum,foss2020holographic}.
The results obtained here will straightforwardly translate to many other quantum 
algorithms for quantum optimization. Examples of such methods include other quantum 
algorithms based on the variational principle \cite{farhi2014quantum,grimsley2019adaptive,mcardle2019variational,barkoutsos2020improving}, quantum algorithms 
for evolution along a prescribed path in the Hilbert space \cite{farhi2000quantum,motta2020determining},
as well as algorithms for long-term quantum devices  \cite{griffiths1996semiclassical,dobvsivcek2007arbitrary,o2019quantum,cruz2020optimizing}.
Our work also enables the exploration of alternative and complementary research directions, 
such as 
the implementation and development of quantum-inspired algorithms for classical computers.

Our work takes a step towards fostering the synergy between quantum computation and numerical simulations for mining applications. While the present work 
offer prospects for quantum algorithms to be applied to the computational design of open pit profiles, it should be regarded to as a proof-of-principle study. Indeed, in view of the formal simplifications and the heuristic algorithms employed here, considerable
methodological and experimental progress is necessary to tackle larger, more challenging and more realistic problem.

A systematic and comparative study of quantum algorithms is required to assess the actual relevancy of 
quantum computation to the present problem.
Further, generalization from two- to three-dimensional systems is also
an important goal, as the dimensionality of the problem has
implications on the required hardware connectivity and the
cost function landscape.
Moreover, improved fragmentation schemes, e.g. allowing for overlapping fragments and systematic choices of the fragmentation choices are also important perspective improvements. 
Finally, penalty-free methods (i.e. algorithms designed to target profiles with $S({\bf{z}})=0$ only, rather than approximately enforcing the smoothness constraint by a penalty operator $\gamma \hat{H}_s$).
Future computational studies can represent an occasion 
for detailed comparison of various computational techniques,
their refinement, and their extension 
to more realistic situations.

\section*{Acknowledgments}

JL and MM acknowledge the IBM Research Cognitive Computing Cluster service for providing 
resources that have contributed to the research results reported within this work
and Barbara Jones, Jennifer Glick, Jeffrey Cohn, and Ryan Mishmash for helpful feedback about this manuscript.
All authors acknowledge support from Stanford University within the CS210 program.
The code used to generate the data presented in this study can be publicly accessed on 
GitHub at \cite{ibmstanford2020gihtub}.

\appendix 

\section{Comparison between optimizers}
\label{sec:app_opt}

In this Section, we compare the performance of different optimizers using the
pit profile in Figure \ref{fig:pit}c as a test case. 
Experiments are run using statevector simulator with penalty $\gamma = 53/3$. 
All the employed optimizers (L-BFGS-B, CG, SLSQP, and SPSA) converge to the
maximum profit of 113, and thus deliver the correct result.
Convergence takes 806, 1333, 797 and 2051 evaluations for L-BFGS-B, CG, 
SLSQP, and SPSA respectively.
Note that multiple evaluations are computed in each optimization iteration. 

\begin{figure}[b!]
\includegraphics[width=0.6\columnwidth]{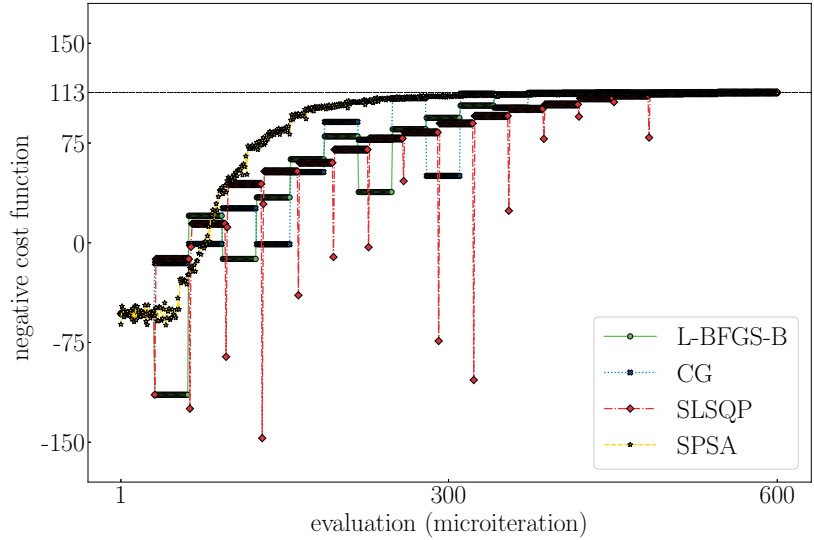}
\caption{Evolution of the negative cost function for the pit profile in Figure 
\ref{fig:pit}c run on a selection of different optimizers with respect to 
evaluation number. 
} 
\label{fig:four_pit_mine_measurement_probability}
\end{figure}

\section{Fragmentation details}
\label{sec:frag}

In this Section, we derive Equations \eqref{eq:frag1} and \eqref{eq:frag2}.
Recall that the unfragmented Hamiltonian for the open-pit mining problem can 
be written as
\begin{equation}
\hat{H} = \sum_i w_i \, \hat{\Pi}^{(1)}_i 
+ \
\sum_{ij} \gamma \, \Delta(ij) \, \hat{\Pi}^{(1)}_i \hat{\Pi}^{(0)}_j
\quad,
\end{equation}
having defined $\hat{\Pi}^{(k)} = |k\rangle \langle k|$ and
\begin{equation}
\Delta(i,j) = 
\left\{
\begin{array}{cc}
1 & \mbox{if $j \in P_i$} \\
0 & \mbox{otherwise} \\
\end{array}
\right.
\end{equation}
for brevity. Introducing a set of fragments and performing the substitution
$\sum_i \to \sum_\alpha \sum_{i\in F_\alpha}$, we readily obtain
\begin{equation}
\hat{H} = \sum_{\alpha} \sum_{i \in F_\alpha} w_i \, \hat{\Pi}^{(1)}_i 
+
\sum_{\alpha} \sum_{ij \in F_\alpha} \gamma \, \Delta(ij) \, \hat{\Pi}^{(1)}_i \hat{\Pi}^{(0)}_j 
+
\sum_{\alpha \neq \beta} \sum_{ \substack{i \in F_\alpha \\ j \in F_\beta} } \gamma \, \Delta(ij) \, \hat{\Pi}^{(1)}_i \hat{\Pi}^{(0)}_j 
\end{equation}
which has the form Equation~\eqref{eq:frag1} with
\begin{equation}
\hat{V}_\alpha 
= 
\sum_{i \in F_\alpha} w_i \, \hat{\Pi}^{(1)}_i + \sum_{ij \in F_\alpha} \gamma \, \Delta(ij)\,  \hat{\Pi}^{(1)}_i \hat{\Pi}^{(0)}_j 
\quad,\quad
\hat{W}_{\alpha\beta} 
= 
\sum_{ \substack{i \in F_\alpha \\ j \in F_\beta} } \gamma \, \Delta(ij) \, \hat{\Pi}^{(1)}_i \hat{\Pi}^{(0)}_j \quad .
\end{equation}
We have thus derived Equation \eqref{eq:frag1}. Let us now derive Equation \eqref{eq:frag2}.
First, we consider the expectation value of $\hat{H}$ over a factored wavefunction,
\begin{equation}
E = \sum_\alpha \langle \Phi_\alpha | \hat{V}_\alpha | \Phi_\alpha \rangle
+ \sum_{\alpha \neq \beta} \langle \Phi_\alpha , \Phi_\beta | \hat{W}_{\alpha,\beta} | \Phi_\alpha , \Phi_\beta \rangle
\end{equation}
To ensure the $\Phi_\alpha$ are normalized, i.e. $\langle \Phi_\alpha | \Phi_\alpha \rangle = 1$
for all $\alpha$, we consider the Lagrangian
$\mathcal{L} = E - \sum_\alpha E_\alpha \langle \Phi_\alpha | \Phi_\alpha \rangle$.
The derivative of $\mathcal{L}$ with respect to $\Phi_\rho^*$ reads
\begin{equation}
\frac{\delta \mathcal{L}}{\delta \Phi_\rho^*} 
= \hat{V}_\rho | \Phi_\rho \rangle 
+ \sum_{\beta \neq \rho} \langle \cdot , \Phi_\beta | \hat{W}_{\rho,\beta} | \Phi_\rho , \Phi_\beta \rangle
+ \sum_{\alpha \neq \rho} \langle \Phi_\alpha  , \cdot | \hat{W}_{\alpha,\rho} | \Phi_\alpha , \Phi_\rho \rangle
- E_\rho | \Phi_\rho \rangle
\end{equation}
and vanishes when
$\hat{T}_\rho | \Phi_\rho \rangle =  E_\rho | \Phi_\rho \rangle$,
with $\hat{T}_\rho$ as in Equation~\eqref{eq:frag2}.

\begin{figure}[b!]
\includegraphics[width=0.12\columnwidth]{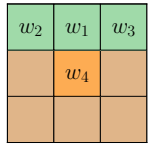}
\caption{Fragmented pit profile.}
\label{fig:frag_prof}
\end{figure}

As an example consider the 4-block mine in Figure~\ref{fig:frag_prof},
partitioned in 2 horizontal fragments (indicated with green and orange).
The upper fragment has the Hamiltonian
\begin{equation}
\begin{split}
    \textrm{H}_1 &= \frac{1}{2}\Big[(w_1 Z_1 + w_2 Z_2 + w_3 Z_3 + w_4 \braket{Z_4}) 
    - (w_1 + w_2 + w_3 + w_4) \Big] \\
    &+ \frac{\gamma}{4} \Big[ (I + Z_1 - \braket{Z_4} - Z_1 \braket{Z_4}) 
    + (I + Z_2 - \braket{Z_4} - Z_2 \braket{Z_4}) 
    + (I + Z_3 - \braket{Z_4} - Z_3 \braket{Z_4}) \Big] 
\end{split}
\end{equation}
and the lower one is described by the Hamiltonian
\begin{equation}
\begin{split}
    \textrm{H}_2 &= \frac{1}{2}\Big[(w_1 \braket{Z_1} + w_2 \braket{Z_2} + w_3 \braket{Z_3} + w_4 Z_4) 
    - (w_1 + w_2 + w_3 + w_4)\Big] \\
    &+ \frac{\gamma}{4} \Big[ (I + \braket{Z_1} - Z_4 - \braket{Z_1} Z_4) 
    + (I + \braket{Z_2} - Z_4 - \braket{Z_2} Z_4) 
    + (I + \braket{Z_3} - Z_4 - \braket{Z_3} Z_4) \Big].
\end{split}
\end{equation}
Thus if a particular pair of blocks contributing to the smoothness Hamiltonian 
is severed by fragmentation, the block outside the fragment of consideration 
still contributes to the smoothness Hamiltonian with its expected value.

\subsection{Comparison between fragmentation strategies}
\label{sec:frag_comparison}

\begin{figure}[t]
\includegraphics[width=0.7\textwidth]{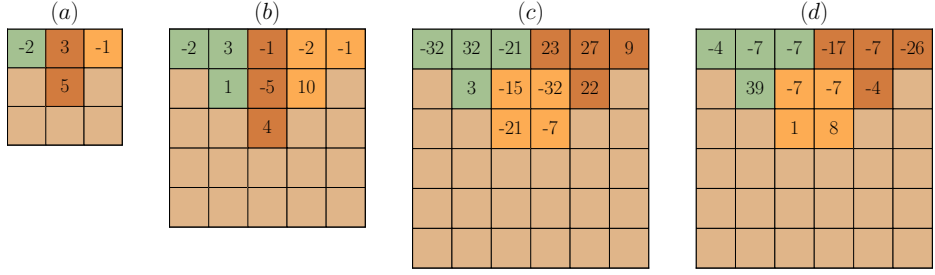}
\caption{\textit{Other} cuts considered in fragmentation simulations.}
\label{fig:pit_o}
\end{figure}

\begin{figure}[t!]
\includegraphics[width=0.8\textwidth]{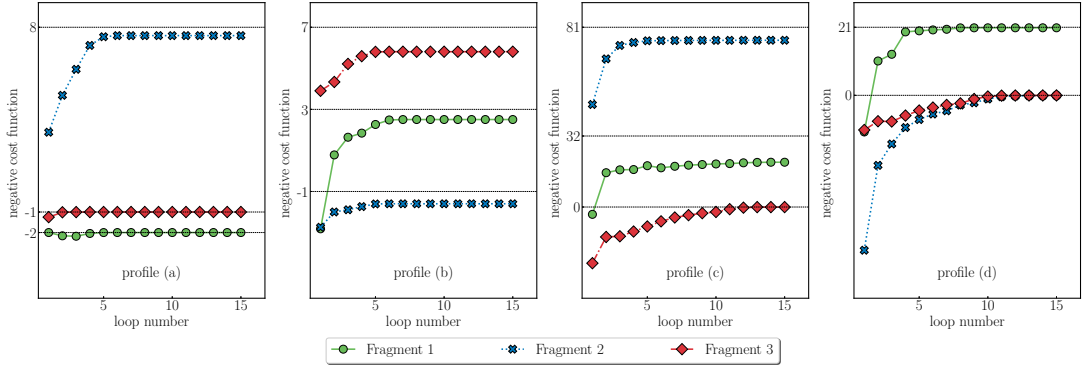}
\caption{
Evolution of the negative cost function of the \textit{other} fragments depicted in Figure \ref{fig:pit_o} using the domain decomposition procedure.}
\label{fig:fragmentation_other}
\end{figure}

\begin{table*}[t!]
\begin{tabular}{ccccccccccccc}
\hline\hline
profile & $|\Psi_0\rangle$ & cut & $\gamma$ & $p_{\mathrm{opt}}$ & optimal & \quad\quad\quad & profile & $|\Psi_0\rangle$ & cut & $\gamma$ & $p_{\mathrm{opt}}$ & optimal \\ 
\hline
a & $| 0 \rangle^{\otimes n}$ & horizontal & 4 & 0.00 & no & \quad & c & $| 0 \rangle^{\otimes n}$ & horizontal & 10 & 0.93 & yes\\
a & $| 1 \rangle^{\otimes n}$ & horizontal & 4 &  \textbf{1.00} & yes & \quad & c & $| 1 \rangle^{\otimes n}$ & horizontal & 10 & 0.92 & yes\\
a & $| + \rangle^{\otimes n}$ & horizontal & 4 &  \textbf{1.00} & yes & \quad & c & $| + \rangle^{\otimes n}$ & horizontal & 10 & \textbf{0.95} & yes\\
a & $| + \rangle^{\otimes n}$ & other & 4 &  0.92 & yes & \quad & c & $| + \rangle^{\otimes n}$ & other & 10 & 0.49 & yes\\
\hline
b & $| 0 \rangle^{\otimes n}$ & horizontal & 5 & 0.00 & no & \quad & d & $| 0 \rangle^{\otimes n}$ & horizontal & 20 & 0.00 & no\\
b & $| 1 \rangle^{\otimes n}$ & horizontal & 5 & 0.00 & no & \quad & d & $| 1 \rangle^{\otimes n}$ & horizontal & 20 & 0.92 & yes\\
b & $| + \rangle^{\otimes n}$ & horizontal & 5 & \textbf{1.00} & yes & \quad & d & $| + \rangle^{\otimes n}$ & horizontal & 20 & 0.92 & yes\\
b & $| + \rangle^{\otimes n}$ & other & 5 & 0.65 & yes & \quad & d & $| + \rangle^{\otimes n}$ & other & 20 & \textbf{0.98} & yes\\\hline\hline
\end{tabular}
\caption{Penalty parameter $\gamma$ and probability 
of detecting an optimal solution $p_{\mathrm{opt}}$,
for the calculations in Figure  \ref{fig:fragmentation_2} and \ref{fig:fragmentation_other}.}
\label{tab:tablefrag}
\end{table*}

There are combinatorially many ways of partitioning a domain into fragments,
and the choice of a fragmentation scheme can drastically affect the quality of final results. 
To assess the sensitivity of VQE calculations to the fragmentation, 
we compare the fragmentation strategies in Figures \ref{fig:pit_h} and \ref{fig:pit_o}.

Results for strategy \ref{fig:pit_h} are shown in the main text, and for 
strategy \ref{fig:pit_o} in Figure \ref{fig:fragmentation_other}.

All simulations are run using statevector simulator with L-BFGS-B optimizer, 
and penalties are set to 4, 5, 10 and 20 respectively. 
Initial states are all set to equal superposition.
In each loop over all fragments, one iteration per fragment is performed. 
All fragments converge to the sum of their excavated sites in the optimal solution, 
thus the sum of the negative cost function of all fragments converges to the maximum 
possible profit of that pit profile, which is 5, 9, 113 and 21 respectively. 

As seen, we obtain the correct solution with ways of cutting the fragments other 
than horizontal fragmentation. The probability of finding the correct bit string is 
generally lower than using horizontal simulations, however the correct bit string 
still has higher probability than all other bit strings in our simulations. 
The reason for the decrease in optimal probability is that when the bond between 
parent-child pairs is preserved within a fragment, we have an additional controlled 
Y-rotation gate preserved in the ansatz circuit, meaning there are more parameters 
to be optimized over. Thus horizontal fragmentation proves to be the most reliable 
but has the disadvantage that it will scale worse than an arbitrary, constant-sized 
way of cutting as the pit size increases. Also note that the equal superposition 
state proves to be the most consistent choice of initial states across different profiles. 
This result is presented in Table \ref{tab:tablefrag}. 
That is the reason why we present fragmentation results with initial states as 
equal superposition and with horizontal cuts in the main section of the present work.
Another technical remark is that during domain decomposition optimization, we 
have experimentally observed that the ansatz parameters can get stuck at the 
boundaries of their allowed range of $[0, \pi]$. If parameters get very close 
to either $0$ or $\pi$, we introduce temperature to the optimization procedure 
by randomly shifting these parameters away from the boundary. For fragments 
where the parent-child pairs have been preserved, we have also observed that 
it is rather beneficial to constrain the sum of the parameters of the single 
and controlled Y-rotation gates acting on a particular qubit to the said 
interval than constraining parameters individually.


%

\end{document}